\documentclass[pdflatex,sn-mathphys-num]{sn-jnl}

\usepackage{graphicx}%
\usepackage{multirow}%
\usepackage{amsmath,amssymb,amsfonts}%
\usepackage{amsthm}%
\usepackage[title]{appendix}%
\usepackage{xcolor}%
\usepackage{textcomp}%
\usepackage{manyfoot}%
\usepackage{booktabs}%
\usepackage{algorithm}%
\usepackage{algorithmicx}%
\usepackage{algpseudocode}%
\usepackage{listings}%
\usepackage{fix-cm}
\usepackage{lmodern}
\usepackage[T1]{fontenc}

\usepackage{physics}%
\usepackage{hyperref}
\usepackage{bbold}

\begin{document}

\title[Article Title]{An Inequality for Entangled Qutrits in SU(3) basis}


\author*[1,2]{\fnm{Surajit} \sur{Sen}}\email{ssen55@yahoo.com}

\author[2]{\fnm{Tushar Kanti} \sur{Dey}}\email{tkdey54@gmail.com}


\affil*[1]{\orgdiv{Physics Department}, \orgname{Guru Charan College}, \\ \orgaddress{ \city{Silchar}, \postcode{788004}, \country{India}}}

\affil[2]{\orgdiv{} \orgname{Centre of Advanced Studies and Innovation Lab}, \\ \orgaddress{\street{18/27 Kali Mohan Road, Tarapur}, \city{Silchar}, \postcode{788003}, \country{India}}}


\abstract{It is well-known from the representation theory of particle physics that the tensor product of two fundamental representation of SU(2) and SU(3) group can be decomposed to obtain the desired spectrum of the physical states. In this paper, we apply this tenet in case of two {\it non-local} qubits and qutrits, which leads the complete spectrum of their entangled states in their respective basis. For qutrit system, the study of their properties reveals the existence of a new $\sqrt{2}$ inequality, in addition to usual Bell-CHSH type $2\sqrt{2}$ inequality, which is significant from the experimental point of view. }




\maketitle

\section{Introduction}\label{sec1}

\par
Quantum entanglement is considered as one of the most exquisite traits of the quantum world \cite{nielsen2000}. During the early years of its development, Einstein-Podolsky-Rosen (EPR) raised their doubt about the ability of quantum theory to explain the nature in a way that is consistent with both the special theory of relativity and the objective realism \cite{epr1935}. Their stance necessitates the postulation of the concept of hidden variables in support of their perspective and also to address some compatibility issues related to the foundational aspects of quantum mechanics. However, soon after that, Bohr refuted their viewpoint by emphasizing on the crucial role of the observer in the measurement process in the quantum realm, which invalidates the proposition of such variables \cite{bohr1935}. Later, the issue became more popular when Bohm reformulated it in a more pragmatic footing  \cite{bohm1951}, enabling Bell \cite{bell1964, bell1993}, and thereafter Clauser-Horne-Shimony-Holt (CHSH) \cite{clauser1969} to point out the existence of an experimentally testable identity. Finally, it was verified experimentally by Aspect and his coworkers \cite{aspect1982}, which was further refined by fixing several loopholes \cite{hensen2015, shalm2015, giustina2015}. Although these studies have laid the foundation of the modern paradigm of quantum information science, the study of entanglement beyond qubit particularly in high dimension, always remains at the center stage for harnessing faster information processing. 
\par 
The simplest representative of an entangled system is the two-qubit system, formally known as the Bell states. This set of entangled states is constructed by the successive operations of the Hadamard and CNOT gates, some variant of the Pauli gates, on the two-qubit state which is intricately linked with the Bell-CHSH inequality \cite{clauser1969,tsirelson1980}. However, in a three-dimensional Hilbert space, the straightforward extension to an entangled qutrit system is not available for various reasons. Recently significant efforts have been made to understand various aspect of qutrits from different perspective. These endeavors include the exploration of the entanglement scenario of the qutrit system 
\cite{caves2000,cereceda2003,rai2005,Bolukbasi2006,jafar2008,pan2006,parsian2019}, the Bloch space structure of qutrits   \cite{arvind1997,kryszewski2006,bengtsson2006,binli2013,goyal2016,Kurzy2016,sharma2021} etc. In addition, since the qutrit system is inherently a three-level system, several quantum-optical phenomena have been investigated within the framework of the SU(3) group \cite{nath2008,sen2012,sen2014,sen2017,sen2023}. Although these studies revealed the crucial role of the SU(3) group to understand various properties of the qutrit system, the information about all conceivable SU(3) based entangled qutrit states is still unknown. The prime hindrance of it is due to the absence of experimentally verifiable inequality for the qutrits, nonavailability of suitable representation of the qutrit gates, and added to them, the complexity arises due to our incomplete understanding of the entanglement scenario of high dimensional Hilbert space.
\par 
There is another reason to study the qutrit system. Soon after the experimental breakthrough of the Bell-CHSH inequality \cite{clauser1969,aspect1982}, it was well understood that an inequality could serve as a powerful tool to verify the non-local character of the quantum correlation. In that context, the qutrit based inequality emerges as a natural choice to address the entanglement scenario of the high dimensional system. In the recent past, the Collins-Gisin-Linden-Massar-Popescu (CGLMP) inequality has garnered significant attention as a multi-partite entangled system, including the qutrit system \cite{collins2002,kasz2002,acin2002,acin2004}. The evaluation of the magnitude of violation from this inequality necessitates the calculation of the expectation value of the CGLMP joint probability operator using an appropriate qutrit state. In this context, the Maximally Entangled State (MES), namely, $\psi_{MES}=\frac{1}{\sqrt{3}}\sum_{j=0}^{2}\ket{j}_A\otimes \ket{j}_{B}$, which is a singlet state of SU(3) group, has been studied in the recent past \cite{cereceda2003,caves2000,collins2002, luo2019,hu2020}. Therefore, it is intriguing to explore other conceivable representations of the two-qutrit states using SU(3) group and then to look for the associated inequality. The measurement setting of the CGLMP operator is determined by various factors such as, the number of parties involved, the number of observable per party, and the number of outcomes per observable \cite{collins2002}. Such requirement imposes a stringent constraint on the geometry of the experimental setup, making it quite difficult for functional implementation. Therefore, it is worthwhile to search for an inequality for the qutrit system in a simpler framework which is closely parallel to the conventional Bell-CHSH setup. 
\par 
The aim of this paper is to generate the complete spectrum of entangled qutrit states within framework of SU(3) group and to find the associated inequality. To accomplish this objective, the paper is structured as follows: In Section II, we develop a heuristic scheme to derive the Bell states from the SU(2) representation theory and review its connection with the Bell-CHSH-Tsirelson bound \cite{clauser1969,tsirelson1980}. In Section III, we extend this methodology to construct all qutrit states using the SU(3) as the basis group and then, discuss their properties in IV. In Section V, we demonstrate how these states can be utilized to develop a pair of inequalities, which we refer as Qutrit Inequality. Finally, we conclude by summarizing the key findings of the paper and discuss the future prospects.

\section{\label{secleve12} SU(2) Group and Bell States}
\par 
To construct the Bell states, let ${q}_i^\alpha$ ($q_i^1=u_i, q_i^2=d_i$) be the fundamental representation of $SU(2)$ group in two-dimensional Hilbert space $\mathcal{H}_i^{2}$ with site index $i$ ($i=A, B$). Following the decomposition of the SU(2) representation of particle physics \cite{Kokkedee1969,greiner1994}, the tensor product of two non-local doublets at site $A$ and $B$ can be expressed as their direct sum, 
\begin{align}\label{1}
\{{\bf {\bar 2}}\}_A\otimes \{{\bf 2}\}_B={\bf 3}_{AB} \oplus {\bf 1}_{AB}, 
\end{align}
where ${\bf 3}_{AB}$ and ${\bf 1}_{AB}$ represent three triplet and one singlet states forming a composite entangled system $AB$ in Hilbert space $\mathcal{H}^{2\otimes2}_{AB}$. 
\par 
To construct Bell states, we adopt a heuristic approach and define the following $\it star $ $\it product$ ($\star $) between the non-local doublets,
\begin{subequations} \label{2}
\begin{align}
{q_A^\alpha}^T \star\sigma_0\star q_B^\alpha &:= {u_A}\star u_B+{d_A}\star d_B,\\
{q_A^\alpha}^T \star\sigma_1\star q_B^\alpha &:= {u_A}\star d_B+{d_A}\star u_B,\\
{q_A^\alpha}^T \star i\sigma_2\star q_B^\alpha &:= {u_A}\star d_B-{d_A}\star u_B,\\
{q_A^\alpha}^T \star\sigma_3\star q_B^\alpha &:= {u_A}\star u_B-{d_A}\star d_B,
\end{align}
\end{subequations}
where $\sigma_i (i=0,1,2,3)$ is the Pauli matrix. Now our method involves two steps: a) promote the star product to tensor product of the basis states ($\star \rightarrow \otimes$), i.e., 
\begin{subequations} \label{3}
\begin{align}
{u_A}\star u_B+{d_A}\star d_B  \rightarrow & \mid \sigma_0\rangle_{AB} := u_A\otimes u_B+{d_A}\otimes d_B,\\
{u_A}\star d_B+{d_A}\star u_B \rightarrow & \mid \sigma_1\rangle_{AB} := u_A\otimes d_B+{d_A}\otimes u_B,\\
{u_A}\star d_B-{d_A}\star u_B \rightarrow & \mid \sigma_2\rangle_{AB} := u_A\otimes d_B-{d_A}\otimes u_B,\\
{u_A}\star u_B-{d_A}\star d_B \rightarrow & \mid \sigma_3\rangle_{AB} := u_A\otimes u_B-{d_A}\otimes d_B,
\end{align}
\end{subequations}
\noindent
and then, b) identify the components of the doublets with the standard qubit basis, namely, $u_{i}:=\ket{0}_i=(1,0)_i^T$, $d_{i}:=\ket{1}_i=(0,1)_i^T$. Thus the normalized Bell states obtained from Eq.\eqref{3} are given by \cite{nielsen2000,bell1993} (For notational convenience, we write $\ket{0}_A \otimes \ket{0}_B = \ket{00}$, $\ket{1}_A \otimes \ket{0}_B = \ket{10}$ etc and drop local indices $i=A,B$),
\begin{subequations} \label{4}
\begin{align}
\ket{\Phi^{+}} &= \frac{1}{\sqrt{2}}\big(\ket{00} + \ket{11}\big),\\
\ket{\Psi^{+}} &=\frac{1}{\sqrt{2}}\big(\ket{01} + \ket{10}\big ),\\
\ket{\Psi^{-}} &=\frac{1}{\sqrt{2}}\big(\ket{01} - \ket{10}\big),\\
\ket{\Phi^{-}} &=\frac{1}{\sqrt{2}}\big(\ket{00} - \ket{11}\big),
\end{align}
\end{subequations}
\noindent
where $\ket{\sigma_0}_{AB} := \ket{\Phi^{+}}$, $\ket{\sigma_1}_{AB} := \ket{\Psi^{+}}$, $\ket{\sigma_2}_{AB} := \ket{\Psi^{-}}$ and $\ket{\sigma_3}_{AB} := \ket{\Phi^{-}}$.
\par
After developing the Bell states, we shall very briefly recall its connection with the Bell-CHSH inequality to understand the subsequent development of the paper. The correlation tensor in $\mathcal{H}^{2\otimes2}_{AB}$ read-off from this inequality is given by \cite{nielsen2000,clauser1969,tsirelson1980},
\begin{align}\label{5}
\hat{\mathcal{C}}^{2\otimes2}=\hat{Q} \otimes \hat{S}+\hat{R} \otimes \hat{S}+\hat{R} \otimes \hat{T}-\hat{Q} \otimes \hat{T},
\end{align}
where $\hat{Q}=\hat{\sigma}_3$, $\hat{S}=-\frac{1}{\sqrt{2}}(\hat{\sigma}_3+\hat{\sigma}_1)$, $\hat{R}=\hat{\sigma}_1$, and $\hat{T}=\frac{1}{\sqrt{2}}(\hat{\sigma}_3-\hat{\sigma}_1)$. This setting can be equivalently expressed as,
\begin{align}\label{6}
\hat{\mathcal{C}}^{2\otimes2}
     =\left(
  \begin{array}{cccc}
    -\sqrt{2} & 0 & 0 & -\sqrt{2} \\
    0 & \sqrt{2} & -\sqrt{2} & 0 \\
    0 & -\sqrt{2} & \sqrt{2} & 0 \\
    -\sqrt{2} & 0 & 0 & -\sqrt{2} \\
  \end{array}\right).  
\end{align}
It is easy to show that the $X$-structured correlation operator is equal to the outer product of two of the Bell states or, equivalently, the tensor product of the Pauli matrices (For detail see, Note added in proof),
\begin{align}\label{7}
\hat{\mathcal{C}}^{2\otimes2}&=
2\sqrt{2}\big(\ket{\Psi^-}\bra{\Psi^-}-\ket{\Phi^+} \bra{\Phi^+}\big) \nonumber \\ 
&=-\sqrt{2}(\hat{\sigma}_3\otimes \hat{\sigma}_3+\hat{\sigma_1}\otimes\hat{\sigma_1}). 
\end{align}
This ensures the non-zero expectation value giving Bell-CHSH-Tsirelson bound  \cite{tsirelson1980,hardy2004},  
\begin{subequations}\label{8}
\begin{align}
\langle \Phi^+ \mid \hat{\mathcal{C}}^{2\otimes2} \mid \Phi^+ \rangle &= -2\sqrt{2},\\
\langle \Psi^- \mid \hat{\mathcal{C}}^{2\otimes2} \mid \Psi^- \rangle &= 2\sqrt{2},
\end{align}
\end{subequations}
while for remaining two Bell states it vanishes. Eq.\eqref{8} is the celebrated bound for the states $\ket{\psi^i}=\{\ket{\Phi^+},\ket{\Psi^-}\}$ known as the Bell-CHSH inequality \cite{nielsen2000}
\begin{equation}\label{9}
{\mid\langle }\psi^i\mid \hat{\mathcal{C}}^{2\otimes2} \mid\psi^i\rangle\mid \leq2\sqrt{2}.
\end{equation}
Thus, it is worth noting that even in the absence of specific information about quantum gates, e.g., Hadamard and CNOT gates, it is still possible to derive the complete spectrum of the entangled states and the inequality associated with it. We now proceed to extend above procedure to derive an inequality for the entangled qutrit system.

\section{\label{secleve13} SU(3) Group and Entangled Qutrit States}
\par 
In three-dimensional Hilbert space $\mathcal{H}^{3}_{i}$ ($i=A$ or $B$), a qutrit system consists of three orthogonal basis states, 
\begin{align}\label{10}
\ket{\psi}=c_0\ket{{0}}+c_1\ket{{1}}+c_2\ket{{2}},
\end{align}
where the standard basis are given by, 
\begin{equation}\label{11}
\begin{split}
\ket{{0}}=\left(
                           \begin{array}{c}
                             1 \\
                             0 \\
                             0 \\
                           \end{array}
                         \right), \quad
\ket{{1}}=\left(
                           \begin{array}{c}
                             0 \\
                             1 \\
                             0 \\
                           \end{array}
                         \right), \quad
\ket{{2}}=\left(
                           \begin{array}{c}
                             0 \\
                             0 \\
                             1 \\
                           \end{array}
                         \right), 
\end{split}
\end{equation}
with the amplitude normalized as $\sum_{i=0}^{2}|c_i|^2=1$. To obtain all possible entangled states in composite Hilbert space $\mathcal{H}^{{3}\otimes{3}}_{AB}$, we first consider the Gell-Mann matrices which in terms of the qutrit basis are given by,
\begin{equation}\label{12}
\begin{aligned}
\lambda_0&=\ket{0}\bra{0}+\ket{1}\bra{1}+\ket{2}\bra{2}, \qquad 
\lambda_1=\ket{0}\bra{1}+\ket{1}\bra{0} \\
\lambda_2&=i(\ket{1}\bra{0}-\ket{0}\bra{1}), 
\qquad  
\lambda_3=\ket{0}\bra{0}-\ket{1}\bra{1}, \\
\lambda_4&=\ket{0}\bra{2}+\ket{2}\bra{0}, \quad 
\qquad 
\lambda_5=i(\ket{2}\bra{0}-\ket{0}\bra{2}), \\ 
\lambda_6&=\ket{1}\bra{2}+\ket{2}\bra{1}, 
\qquad 
\lambda_7=i(\ket{2}\bra{1}-\ket{1}\bra{2}), \\ 
\lambda_8 & =\frac{1}{\sqrt{3}}(\ket{0}\bra{0}+\ket{1}\bra{1}-2\ket{2}\bra{2}).
\end{aligned}
\end{equation}
\noindent
Here the $\lambda_i$ matrices are normalized as $\lambda_l\lambda_m=\delta_{lm}+d_{lmn}\lambda_n+f_{lmp}\lambda_p$ with $d_{lmn}$ and $f_{lmp}$ ($l,m,n,p=1,2, \dots,8$) as the completely symmetric and completely antisymmetric structure constants \cite{greiner1994}. Similar to the previous section, we now consider two non-local triplet $q_i^{\alpha}=(u_i, d_i, s_i)^T \in \mathcal{H}_i^{3}$ and promote the {\it star product} into tensor product,
\begin{subequations}\label{13}
\begin{align}
{q_A^\alpha}^T\star \lambda_0 \star {q_B^\alpha} \rightarrow & \mid \lambda_0\rangle_{AB}:= u_A\otimes u_B+d_A\otimes d_B+s_A\otimes s_B,\\
{q_A^\alpha}^T\star \lambda_1 \star {q_B^\alpha} \rightarrow & \mid \lambda_1\rangle_{AB}:= s_A\otimes v_B+v_B\otimes s_A,\\
{q_A^\alpha}^T\star i\lambda_2 \star {q_B^\alpha} \rightarrow & \mid \lambda_2\rangle_{AB}:= -s_A\otimes v_B+v_B\otimes s_A,\\
{q_A^\alpha}^T\star \lambda_3 \star {q_B^\alpha} \rightarrow & \mid \lambda_3\rangle_{AB}:= -v_A\otimes v_B+s_A\otimes s_B,\\
{q_A^\alpha}^T\star \lambda_4 \star {q_B^\alpha} \rightarrow & \mid \lambda_4\rangle_{AB}:= u_A\otimes s_B+s_B\otimes u_A,\\
{q_A^\alpha}^T\star i\lambda_5 \star {q_B^\alpha} \rightarrow & \mid \lambda_5\rangle_{AB}:= -u_A\otimes s_B+s_B\otimes u_A,\\
{q_A^\alpha}^T\star \lambda_6 \star {q_B^\alpha} \rightarrow & \mid \lambda_6\rangle_{AB}:= u_A\otimes v_B+v_B\otimes u_A,\\
{q_A^\alpha}^T\star i\lambda_7 \star {q_B^\alpha} \rightarrow & \mid \lambda_7\rangle_{AB}:= -u_A\otimes v_B+v_B\otimes u_A,\\
{q_A^\alpha}^T\star \lambda_8 \star {q_B^\alpha} \rightarrow & \mid \lambda_8\rangle_{AB}:= -2 u_A\otimes u_B+d_A\otimes d_B+s_A\otimes s_B. 
\end{align}
\end{subequations}
\noindent
If we identify the states $\{u_{i}, d_{i}, s_{i}\}$ with the standard basis in Eq.\eqref{12}, i.e., $u_{i}=\ket{0}_i$, $d_{i}=\ket{1}_i$, $s_{i}=\ket{2}_i$, Eq.\eqref{13} leads to the normalized entangled qutrit states, 
\begin{subequations}\label{14}
\begin{align}
\ket{\psi_{00}}&=\frac{1}{\sqrt{3}}\big(\ket{00} + \ket{11} + \ket{22} \big), \\
\ket{\psi_{21}^+}&=\frac{1}{\sqrt{2}}\big(\ket{21} + \ket{12} \big), \\ 
\ket{\psi_{21}^-}&=\frac{1}{\sqrt{2}}\big(\ket{21} - \ket{12} \big), \\
\ket{\psi_{11}}&=\frac{1}{\sqrt{2}}\big(-\ket{11} + \ket{22}\big),\\ 
\ket{\psi_{20}^+}&=\frac{1}{\sqrt{2}}\big(\ket{20} + \ket{02} \big)\\
\ket{\psi_{20}^-}&=\frac{1}{\sqrt{2}}\big(\ket{20} - \ket{02} \big),\\
\ket{\psi_{10}^+}&=\frac{1}{\sqrt{2}}\big(\ket{10} + \ket{01} \big), \\
\ket{\psi_{10}^-}&=\frac{1}{\sqrt{2}}\big(\ket{10} - \ket{01} \big),\\
\ket{\psi_{22}}&=\frac{1}{\sqrt{6}}\big(-2\ket{00} + \ket{11} + \ket{22} \big).
\end{align}
\end{subequations}
\noindent 
where $\ket{\lambda_0}_{AB} :=  \ket{\psi_{00}}$, $\ket{\lambda_1}_{AB} :=  \ket{\psi_{21}^{+}}$ etc. Thus we note that, analogous to $SU(2)$ theory, two qutrits follow SU(3) decomposition, namely, $\{{\bf {\bar 3}}\}_A\otimes \{{\bf {3}}\}_B={\bf 8}_{AB}\oplus {\bf 1}_{AB}$ in Hilbert space $\mathcal{H}^{3\otimes3}_{AB}$ which leads to eight octet-like and one singlet-like entangled states. Among them, only the singlet state $\ket{\psi_{00}}$ has been recently investigated both theoretically \cite{cereceda2003,collins2002,acin2002} and experimentally \cite{luo2019,hu2020}, while other states remain unexplored. 

\section{\label{sec:level5} Properties of Entangled Qutrit States}
\par 
Before delivering the inequalities associated with these states, we shall touch upon their properties. We note that, on swapping positions, out of the nine states, $\ket{\psi_{00}}, \ket{\psi_{10}^+}, \ket{\psi_{20}^+}, \ket{\psi_{21}^+}$ are symmetric states, $\ket{\psi_{11}}, \ket{\psi_{10}^-}, \ket{\psi_{20}^-}, \ket{\psi_{21}^-}$
are antisymmetric states, while $\ket{\psi_{22}}$ does not conform to this pattern. 
\par 
To see whether their subsystems are pure or mixed state, it is customary to consider the reduced density matrix of the system. Taking the partial trace over one subsystem (say, B-subsystem) we obtain, 
\begin{align}\label{15}
\rho_{00}^{A}&=\frac{1}{3}\big(\ket{0}_A\bra{0}+\ket{1}_A\bra{1}+\ket{2}_A\bra{2}\big), \nonumber\\ 
{\rho_{10}^{+}}^A &={\rho_{10}^{-}}^A=\frac{1}{2}(\ket{0}_A\bra{1}+\ket{1}_A\bra{0}\big), \\ 
{\rho_{20}^{+}}^A&={\rho_{20}^{-}}^A=\frac{1}{2}\big(\ket{0}_A\bra{2}+\ket{2}_A\bra{0}\big), \nonumber \\ 
{\rho_{21}^{+}}^A&={\rho_{21}^{-}}^A=\rho_{11}^{A}=\frac{1}{2}\big(\ket{1}_A\bra{2}+\ket{2}_A\bra{1}\big), \nonumber \\ 
\rho_{22}^{A}&=\frac{1}{6}\big(4\ket{0}_A\bra{0}+\ket{1}_A\bra{1}+\ket{2}_A\bra{2}\big), \nonumber 
\end{align}
We note that the reduced density matrices satisfy the properties  $Tr[\rho^{A}] = 1$ and $Tr[{\rho^{A}}^{2}] < 1$, indicating that they are essentially a mixed state. The maximally mixed state can be corroborated by noting the entropy,  $S^A_{00}=\log_23 := 1.585$ for the singlet state in comparison to the other states, $S^A_{ij}=1$ and $S^A_{22}=1.2516$. 
\par
For completeness, we finally discuss the change of basis of the bipartite qutrit system. In the computational basis, most general entangled two-qutrit system is given by, 
\begin{equation}\label{16}
\begin{aligned}
\ket{\Psi_T} &=c_{00}\ket{00}+c_{01}\ket{01}+c_{02}\ket{02} \\
&+c_{10}\ket{10} + c_{11}\ket{11} +c_{12}\ket{12} \\
&+c_{20}\ket{20}+c_{21}\ket{21} + c_{22}\ket{22}.
\end{aligned}
\end{equation}
where the amplitudes are normalized as $\sum_{i,j}|c_{ij}|^2=1$. Plucking back $\ket{00}$, $\ket{01}$ etc from Eq.\eqref{14} into Eq.\eqref{16}, the wave function can be transformed into $SU(3)$ basis,
\begin{equation}\label{17}
\begin{aligned}
\ket{\Psi_{SU(3)}} &= b_{00}\ket{\psi_{00}}+b_{21}^+\ket{\psi_{21}^+}+b_{21}^-\ket{\psi_{21}^-}, \\
&+b_{11}\ket{\psi_{11}}+b_{20}^+\ket{\psi_{20}^+}+b_{20}^-\ket{\psi_{02}^-}, \\
&+b_{10}^+\ket{\psi_{10}^+}+b_{10}^-\ket{\psi_{10}^-}+b_{22}\ket{\psi_{22}},
\end{aligned}
\end{equation}
where the amplitudes are given by,
\begin{equation}\label{18}
\begin{aligned}
b_{00}=&\frac{1}{\sqrt{3}}(c_{00}-\sqrt{2}c_{22}), \quad b_{21}^+=\frac{1}{\sqrt{2}}({c_{01}}+{c_{02}}), \quad b_{21}^-=\frac{1}{\sqrt{2}}({c_{01}}-{c_{02}}),\\
b_{11}=&\frac{1}{\sqrt{6}}(\sqrt{2}c_{20}-\sqrt{3}c_{02}+c_{22}), \quad b_{20}^+=\frac{1}{\sqrt{2}}({c_{20}}+{c_{21}}), \quad b_{20}^-=\frac{1}{\sqrt{2}}({c_{20}}-{c_{21}}), \\
b_{10}^+=&\frac{1}{\sqrt{2}}({c_{11}}+{c_{12}}), \quad b_{10}^-=\frac{1}{\sqrt{2}}({c_{11}}-{c_{12}}), \quad b_{22}=\frac{1}{\sqrt{6}}(\sqrt{2}c_{00}+\sqrt{3}c_{10}+c_{22}). 
\end{aligned}
\end{equation}
In this new basis, the amplitudes once again are normalized, i.e.,  $\sum_{i,j}|b_{ij}|^2=1$, indicating the consistency of our construction of all entangled qutrit states. In both basis, the density matrix satisfies the pure state condition, $\rho^2_T=\rho_T$ and $\rho^2_{SU(3)}=\rho_{SU(3)}$. 

\section{\label{secleve16} Qutrit Inequality}
\par
To this end, we are finally in position to discuss the inequality associated with the states obtained above. To construct the correlation tensor using SU(3) as the basis group, we adopt the methodology developed above. Similar to Eq.\eqref{6}, the correlation tensor for the qutrit system should satisfy the following properties: i) it must be a symmetric matrix, 
ii) it should be diagonal in the qutrit basis states $\ket{\psi_{ij}}$ and, iii) it should be expressed as the tensor product of the $SU(3)$ matrices \cite{hardy2004}. After a lengthy but straight forward calculation yields, 
\begin{align}\label{19}
\hat{\mathcal{C}}^{3\otimes3} &= \left(
        \begin{array}{ccccccccc}
          0 & 0 & 0 & 0 & \sqrt{2} & 0 & 0 & 0 & \sqrt{2} \\
          0 & \sqrt{2} & 0 & -\sqrt{2} & 0 & 0 & 0 & 0 & 0 \\
          0 & 0 & -\sqrt{2} & 0 & 0 & 0 & \sqrt{2} & 0 & 0 \\
          0 & -\sqrt{2} & 0 & \sqrt{2} & 0 & 0 & 0 & 0 & 0 \\
          \sqrt{2} & 0 & 0 & 0 & \sqrt{2} & 0 & 0 & 0 & 0 \\
          0 & 0 & 0 & 0 & 0 & -\sqrt{2} & 0 & \sqrt{2} & 0 \\
          0 & 0 & \sqrt{2} & 0 & 0 & 0 & -\sqrt{2} & 0 & 0 \\
          0 & 0 & 0 & 0 & 0 & \sqrt{2} & 0 & -\sqrt{2} & 0 \\
          \sqrt{2} & 0 & 0 & 0 & 0 & 0 & 0 & 0 & \sqrt{2} \\
        \end{array}
      \right). 
\end{align}
It is worth mentioning here that, similar to Bell system, the diagonalization of above correlation matrix precisely gives the Bell-like qutrit states Eq.\eqref{14} as its eigenvectors \cite{hardy2004}. Finally it is easy to see that the matrix can be decomposed in terms of their outer product (See, Note added in proof), 
\begin{align}\label{20}
\hat{\mathcal{C}}^{3\otimes3}
&= \sqrt{2}\big(\ket{\psi_{11}}\bra{\psi_{11}}-\ket{\psi_{22}}\bra{\psi_{22}}\big) \nonumber \\
&+2\sqrt{2}\big(\ket{\psi_{00}}\bra{\psi_{00}} 
+\ket{\psi_{10}^-}\bra{\psi_{10}^-}-\ket{\psi_{12}^-}\bra{\psi_{12}^-}-\ket{\psi_{20}^-}\bra{\psi_{20}^-}\big)  
\end{align}
which can be further expressed as the tensor product of GellMann matrices, 
\begin{align}\label{21}
\hat{\mathcal{C}}^{3\otimes3}
&=\sqrt{2}(\lambda_4\otimes\lambda_4-\lambda_2\otimes\lambda_2)+\frac{1}{\sqrt{2}}(\lambda_6\otimes\lambda_6  + \lambda_7\otimes\lambda_7) \nonumber \\
&-\frac{1}{2\sqrt{2}}(\lambda_3\otimes\lambda_3-5\lambda_8\otimes\lambda_8) +\frac{1}{\sqrt{6}}(\lambda_8\otimes\lambda_0+\lambda_0\otimes\lambda_8) \\ 
&-\frac{1}{2\sqrt{6}}(\lambda_3\otimes\lambda_8+\lambda_8\otimes\lambda_3)-\frac{1}{3\sqrt{2}}(\lambda_0\otimes\lambda_3 +\lambda_3\otimes\lambda_0). \nonumber
\end{align}
The basis in which the correlation matrix is diagonal leads to the following bounds, 
\begin{equation}\label{22}
\begin{aligned}
\langle\psi_{ij} \mid  \hat{\mathcal{C}}^{3\otimes3} \mid \psi_{ij} \rangle &= \sqrt{2} \quad \textrm{for} \quad \{\ket{\psi_{11}}\},\\
\langle\psi_{ij} \mid \hat{\mathcal{C}}^{3\otimes3} \mid \psi_{ij} \rangle &= 2\sqrt{2} \quad \textrm{for} \quad \{\ket{\psi_{00}}, \ket{\psi_{10}^-}\},\\
\langle\psi_{ij} \mid \hat{\mathcal{C}}^{3\otimes3} \mid \psi_{ij} \rangle &= -\sqrt{2} \quad \textrm{for} \quad \ket{\{\psi_{22}}\},\\
\langle\psi_{ij} \mid \hat{\mathcal{C}}^{3\otimes3} \mid \psi_{ij} \rangle &= -2\sqrt{2} \quad \textrm{for} \quad \{\ket{\psi_{12}^-},\ket{\psi_{20}^-}\}.  
\end{aligned}
\end{equation}
while for the remaining states, namely, $\{\ket{\psi_{12}^+}, \ket{\psi_{20}^+}, \ket{\psi_{10}^+}\}$, it vanishes. Thus we have two distinct set of inequalities for the entangled qutrit system, 
\begin{equation}\label{23}
\begin{aligned}
\mid\langle\psi_{ij} \mid \hat{\mathcal{C}}^{3\otimes3} \mid \psi_{ij} \rangle \mid & \leq 2\sqrt{2}  \quad \textrm{for} \quad \{\ket{\psi_{00}}, \ket{\psi_{12}^-}, \ket{\psi_{20}^-}, \ket{\psi_{20}^-}\}, \\
\mid\langle\psi_{ij} \mid \hat{\mathcal{C}}^{3\otimes3} \mid \psi_{ij} \rangle \mid & \leq \sqrt{2},  \quad \textrm{for} \quad \{\ket{\psi_{11}}, \ket{\psi_{22}}\}.  
\end{aligned}
\end{equation}
These are desired Qutrit Inequality for the qutrit system. The emergence of the Bell-CHSH-like $2\sqrt{2}$ inequality in the sideline of our treatment shows the consistency of the SU(3) based approach of which SU(2) is an subgroup.  

\section{\label{secleve17} Conclusion}
\par
In this paper, using the representation theory of the SU(2) and SU(3) group, we present a method to derive the Bell states for qubit and Bell-like entangled states for qutrit systems. Our approach, which is inspired by particle physics, uses a heuristic {\it star product} method to generate the entire spectrum of entangled states without prior information about the quantum gates. After discussing the key properties of these states and their special attributes, we proceed to find the correlation function of  such system which can be expressed in terms of the tensor product of the Gell-Mann matrices. It is noteworthy that the appearance of a new $\sqrt{2}$ inequality, apart from the well-known Bell-CHSH $2\sqrt{2}$ inequality, without prior knowledge of the detector setting, is a natural outcome of our analysis, which may have experimental importance. In the ever-expanding landscape of quantum technology, the qutrits may play a significant role in shaping the future quantum computers with improved algorithms, augmented communication ability and multifaceted security.
\vfill 

\section*{Acknowledgement}
\par 
We thank Professor Bijan Bagchi for fruitful discussion.

\pagebreak 
\section*{\label{secleve1Note} Note added in proof:}

Here we give an outline of the derivation of Eq.\eqref{7} and Eq.\eqref{21} from the entangled states of the Bell and Bell-like Qutrit systems, which is crucial to understand the derivation of the corresponding inequalities. 
\par 
To derive the correlation matrix for the Bell states, we start with the linear combination of all possible density matrix operators obtained by taking the outer product of the Bell states, 
\begin{align}\label{24}
& a_1\rho_{\Phi^+} + a_2\rho_{\Phi^-}+a_3\rho_{\Psi^+} +a_4\rho_{\Psi^-} \nonumber \\ 
=&a_1\ket{\Phi^+}\bra{\Phi^+} + a_2\ket{\Phi^-}\bra{\Phi^-}+a_3\ket{\Psi^+}\bra{\Psi^+} +a_4\ket{\Psi^-}\bra{\Psi^-} \nonumber \\ 
&=\begin{pmatrix}
     \frac{1}{2}(a_1+a_2) & 0 & 0 & \frac{1}{2}(a_1-a_2)\\ 
     0 & \frac{1}{2}(a_3+a_4) & \frac{1}{2}(a_3-a_4) & 0\\
     0 & \frac{1}{2}(a_3-a_4) & \frac{1}{2}(a_3+a_4) & 0\\
     \frac{1}{2}(a_1-a_2) & 0 & 0 & \frac{1}{2}(a_1+a_2)
 \end{pmatrix}, 
\end{align}
where $a_i$ ($i= 1, 2,3, 4$) be constants to be determined. Comparing Eq.\eqref{24} with Eq.\eqref{6}, the constants are found to be $a_1=-2\sqrt{2}$, $a_2=a_3=0$, $a_4=2\sqrt{2}$ and we have, 
\begin{align}\label{25}
& a_1\rho_{\Phi^+} + a_2\rho_{\Phi^-}+a_3\rho_{\Psi^+} +a_4\rho_{\Psi^-} \nonumber \\ 
&=2\sqrt{2}\big(\ket{\Psi^-}\bra{\Psi^-}-\ket{\Phi^+} \bra{\Phi^+}\big)  \nonumber \\ 
&=-\sqrt{2}(\hat{\sigma}_3\otimes \hat{\sigma}_3+\hat{\sigma_1}\otimes\hat{\sigma_1}) \nonumber \\
&=\hat{\mathcal{C}}^{2\otimes2} \quad Q.E.D. 
\end{align}
In the last step we used the fact that the outer product of the Bells states is numerically equal to the tensor product of the Pauli matrices.
\par 
To derive the correlation tensor for the two-qutrit system, similar to the previous case, we consider the linear combination of the density matrices from all nine two-qutrit states derived in Eq.\eqref{14},  
\begin{align}\label{27}
a_{00}\rho_{\psi_{00}}&+a_{12}^+\rho_{\psi_{12}^+}+a_{12}^-\rho_{\psi_{12}^-} +  a_{11}\rho_{\psi_{11}}  \nonumber \\ 
&+ a_{20}^+\rho_{\psi_{20}^+} + a_{20}^-\rho_{\psi_{20}^-} +a_{10}^+
\rho_{\psi_{10}^+} + a_{10}^-\rho_{\psi_{10}^-}+a_{22}\rho_{\psi_{22}} \nonumber \\
&=a_{00}\ket{\psi_{00}}\bra{\psi_{00}}+a_{12}^+\ket{\psi_{12}^+}\bra{\psi_{12}^+}+
a_{12}^-\ket{\psi_{12}^-}\bra{\psi_{12}^-}  \nonumber \\
&+a_{11}\ket{\psi_{11}}\bra{\psi_{11}}
+a_{20}^+\ket{\psi_{20}^+}\bra{\psi_{20}^+}+a_{20}^-\ket{\psi_{20}^-}\bra{\psi_{20}^-}  \nonumber \\
&+ a_{10}^+\ket{\psi_{10}^+}\bra{\psi_{10}^+} + a_{10}^-\ket{\psi_{10}^-}\bra{\psi_{10}^-} 
+a_{22}\ket{\psi_{22}}\bra{\psi_{22}} 
\end{align}
In absence of the correlation matrix for the qutrit system, aforesaid procedure of evaluating constants by simple comparison does not work. We therefore adopt a procedure  where the diagonalization of that correlation matrix must give all two-qutrit states as its eigen states \cite{hardy2004}. This requirement, supplemented by rigorous algebraic manipulation of a set of coupled linear equations gives, $a_{00}=2\sqrt{2}$, $a_{12}^+=0$, $a_{12}^-=-2\sqrt{2}$, $a_{11}=\sqrt{2}$, $a_{20}^+=0$, $a_{20}^-=-2\sqrt{2}$, $a_{10}^+=0$, $a_{10}^-=2\sqrt{2}$, $a_{22}=-\sqrt{2}$.
Plugging them back in Eq.\eqref{27}, we obtain the qutrit correlation matrix given by Eq.\eqref{20}, which can be also written as the tensor product of the GellMann matrices,
\begin{align}\label{26}
&a_{00}\rho_{\psi_{00}}+a_{12}^+\rho_{\psi_{12}^+}+a_{12}^-\rho_{\psi_{12}^-} +  a_{11}\rho_{\psi_{11}}  \nonumber \\ 
&+ a_{20}^+\rho_{\psi_{20}^+} + a_{20}^-\rho_{\psi_{20}^-}+a_{10}^+ 
\rho_{\psi_{10}^+} + a_{10}^-\rho_{\psi_{10}^-}+a_{22}\rho_{\psi_{22}}
\\
&= \sqrt{2}\big(\ket{\psi_{11}}\bra{\psi_{11}}-\ket{\psi_{22}}\bra{\psi_{22}}\big) \nonumber \\
&+2\sqrt{2}\big(\ket{\psi_{00}}\bra{\psi_{00}} 
+\ket{\psi_{10}^-}\bra{\psi_{10}^-}-\ket{\psi_{12}^-}\bra{\psi_{12}^-}-\ket{\psi_{20}^-}\bra{\psi_{20}^-}\big) \nonumber \\
&=\sqrt{2}(\lambda_4\otimes\lambda_4-\lambda_2\otimes\lambda_2) 
+\frac{1}{\sqrt{2}}(\lambda_6\otimes\lambda_6  + \lambda_7\otimes\lambda_7) \nonumber \\
&-\frac{1}{2\sqrt{2}}(\lambda_3\otimes\lambda_3-5\lambda_8\otimes\lambda_8) 
+\frac{1}{\sqrt{6}}(\lambda_8\otimes\lambda_0+\lambda_0\otimes\lambda_8) \nonumber \\ 
& -\frac{1}{2\sqrt{6}}(\lambda_3\otimes\lambda_8+\lambda_8\otimes\lambda_3)-\frac{1}{2\sqrt{3}}(\lambda_0\otimes\lambda_3 +\lambda_3\otimes\lambda_0),  \nonumber \\
&=\hat{\mathcal{C}}^{3\otimes3} \qquad \qquad Q.E.D.  \nonumber  \tag{28}
\end{align}

\vfill


\pagebreak 


\pagebreak 

\begin{thebibliography}{50}
\bibitem{nielsen2000} \hypertarget{nielsen2000}
M.A. Nielsen and I.L. Chuang, {\it Quantum Information and Quantum Computation}, Cambridge University Press, Cambridge, (2000)
\bibitem{epr1935}
A. Einstein, B. Podolsky and N. Rosen: Can Quantum Mechanical Description of Physical Reality Be Considered Complete? Phys. \ Rev. {\bf 47}, 777 (1935)
\bibitem{bohr1935}
N. Bohr: Can quantum-mechanical description of physical reality be considered complete?, Phys. Rev. {\bf 48} 696–702 (1935)
\bibitem{bohm1951}
D. Bohm, {\it Quantum Theory}, Prentice Hall, New York, (1951)
\bibitem{bell1964}
J. S. Bell: On the Einstein-Podolsky-Rosen paradox Physics {\bf 1}, 195-200 (1964)
\bibitem{bell1993}
Ref[4] reprinted in, J. S. Bell, Speakable and unspeakable in quantum mechanics : collected papers on quantum philosophy, Cambridge University Press, Cambridge (1993)
\bibitem{clauser1969}
J.F. Clauser, M.A. Horne, A. Shimony and R.A. Holt: Proposed Experiment to Test Local Hidden-Variable Theories. Phys. Rev. Lett. {\bf 23}, 880-4 (1969)
\bibitem{tsirelson1980}
B. S. Cirel'son: Quantum generalizations of Bell's inequality, Letters in Mathematical Physics, {\bf 4}, 93 (1980)
\bibitem{aspect1982}
A. Aspect, P. Grangier and G. Roger: Experimental Realization of Einstein-Podolsky-Rosen-Bohm Gedankenexperiment: A New Violation of Bell's Inequalities. Phys. Rev. Lett. {\bf 49}, 91 (1982)
\bibitem{hensen2015}
B. Hensen el al.: Loophole-free Bell inequality violation using electron spins separated by 1.3 kilometres. Nature {\bf 526}, 682--686 (2015)
\bibitem{shalm2015}
L.K. Shalm et al.: Strong Loophole-Free Test of Local Realism. Phys. Rev Lett. {\bf 115}, 250402 (2015)
\bibitem{giustina2015}
M. Giustina et. al.; Significant-Loophole-Free Test of Bell’s Theorem with Entangled Photons. Phys. Rev. Lett. {\bf 115}, 250401 (2015)
\bibitem{caves2000}
C. M. Caves and G. J. Milburn: Qutrit Entanglement, Optics Comm. {\bf 179} 439-446 (2000)
\bibitem{cereceda2003}
J. L. Cereceda: Degree of entanglement for two qutrits in a pure state. quant/ph: 0305043
\bibitem{rai2005}
S. Rai and J. R. Luthra: Negativity and Concurrence for two qutrits. quant-ph/: 0507263
\bibitem{Bolukbasi2006}
B\"ol\"ukbasi, A. T. and Dereli, T.: On the SU(3) Parametrization of Qutrits, Jour. of Phys.: Conference Series. {\bf 36} 28-32 (2006)
\bibitem{jafar2008}
M. A. Jafarizadeh, Y. Akbari and N. Behzadi: Two-qutrit entanglement witnesses and Gell-Mann matrices. The Euro. Phys. Jour. {\bf D47} 283–293 (2008)
\bibitem{pan2006}
F. Pan and G. Lu: Classification and Quantification of Entangled Bipartite Qutrit Pure States. Int. J. Mod. Phys. {\bf 20} 1333-1342 (2006) 
\bibitem{parsian2019}
H. Parsian and A. Akhound: Classical and quantum correlations for a family of two-qutrit states. Int. J. Quant. Inf. {\bf 17}, 1950028 (2019)
\bibitem{arvind1997}
Arvind, K. S. Mallesh and N. Mukunda: A generalized Pancharatnam geometric phase formula for three level quantum systems, J. Phys. A. Math. Gen. {\bf 30} 2417 (1997)
\bibitem{kryszewski2006}
S. Kryszewski, M. Zachcial: Alternative representation of $N \otimes N$ density matrix, quant-ph/0602065
\bibitem{bengtsson2006}
I.Bengtsson and K.Życzkowski, {\it Geometry of Quantum States} Cambridge University Press, Cambridge, (2006)
\bibitem{binli2013}
L. Bin, Y. Zu-Huan and F. Shao-Ming: Geometry of Quantum Computation with
Qutrits. Sci. Rep. {\bf 3}, 2594 (2013).
\bibitem{goyal2016}
S. Goyal, B.N. Simon, R. Singh, and S. Simon: Geometry of the generalized Bloch sphere for qutrits, J. Phys. A: Math. Theor. {\bf 49} 165203 (2016)
\bibitem{Kurzy2016}
P. Kurzyński, A. Kolodziejski, W. Laskowski and M. Markiewicz: 
Three-dimensional visualization of a qutrit. Phys. Rev. A {\bf 93} 062126 (2016)
\bibitem{sharma2021}
G. Sharma and S. Ghosh, Four-dimensional Bloch sphere representation of qutrits using Heisenberg-Weyl Operators, quant-ph/: 2101.06408v2
\bibitem{sen2012}
S. Sen, M. R. Nath, T. K. Dey and G. Gangopadhyay: Bloch space structure, the qutrit wave function and atom–field entanglement in three-level systems. Ann. Phys. {\bf 327}, 224--252 (2012)

\bibitem{nath2008}
M. R. Nath, S. Sen, A. K. Sen and G. Gangopadhyay; Dynamical symmetry breaking of lambda and vee-type three-level systems on quantization of the field modes, Pramana - Jour. of Phys. {\bf 71}, 77--97 (2008) 

\bibitem{sen2017}
S. Sen,T. K. Dey and B. Deb; A unified approach to $\Lambda$, $V$ and $\Xi$-type systems with one continuum, Jour of Mod. Opt. {\bf 64} 2083--2096 (2017) 
\bibitem{sen2023}
S. Sen,T. K. Dey and B. Deb; Resonance fluorescence in $\Lambda$, $V$ and $\Xi$-type three-level configurations, Phys. Scr. {\bf 98} 115124 (2023) 
\bibitem{sen2014}
S. Sen, T K Dey, M. R. Nath and G. Gangopadhyay; Comparison of Electromagnetically Induced Transparency in lambda, cascade and vee three-level systems, Jour. of Mod. Opt. {\bf 62} 166--174 (2014) 
\bibitem{collins2002}
D. Collins, N. Gisin, N. Linden, S. Massar, S. Popescu: Bell Inequalities for Arbitrarily High-Dimensional Systems, Phys. Rev. Lett. {\bf 88}, 040404 (2002)
\bibitem{kasz2002}
D. Kaszlikowski, L. C. Kwek, J. L. Chen, M. Zukowski, and C. H. Oh: Clauser-Horne inequality for three-state systems. Phys. Rev. {\bf A65} 032118 (2002)
\bibitem{acin2002}
A. Acín, T. Durt, N. Gisin, and J. I. Latorre: Quantum nonlocality in two three-level systems, Phys. Rev. {\bf A65}, 052325 (2002)
\bibitem{acin2004}
A. Acín, J. L. Chen, N. Gisin, D. Kaszlikowski, L. C. Kwek, C. H. Oh and M. Zukowski: Coincidence bell inequality for three three-dimensional systems. Phys. Rev. Lett. {\bf 92} 250404 (2004)
\bibitem{luo2019} 
Yi-Han Luo et. al.: Quantum Teleportation in High Dimensions. Phys. Rev. Lett. {\bf 123} 070505 (2019)
\bibitem{hu2020} 
X.M. Hu et. al.: Experimental High-Dimensional Quantum Teleportation Phys. Rev. Lett. {\bf 125}, 230501 (2020)
\bibitem{Kokkedee1969}
For example, see, J. J. J. Kokkedee, {\it The Quark Model} W. A. Benjamin Inc., New York, (1969)
\bibitem{greiner1994}
See also, W. Greiner and B. M$\ddot{u}$ller, {\it Quantum Mechanics: Symmetries} Springer Verlag, Berlin, (1994)
\bibitem{hardy2004}
W. H. Steeb and Y. Hardy, Problems and Solutions in Quantum Computing and Quantum Information, p.112, World Scientific, (2004)
\end{thebibliography}

\end{document}